\documentclass[runningheads]{llncs}

\usepackage{longtable}
\usepackage{philex}
\usepackage{amssymb}
\usepackage{bussproofs}

\usepackage{stackengine}
\stackMath

\newtheorem{manualtheoreminner}{Theorem}
\newenvironment{manualtheorem}[1]{%
  \renewcommand\themanualtheoreminner{#1}%
  \manualtheoreminner
}{\endmanualtheoreminner}

\usepackage{tikz}
\usepackage{tikz-qtree}

\begin{document}

\title{Proof-Theory and Semantics for a Theory of Definite Descriptions\thanks{The research in this paper was funded by the Alexander von Humboldt Foundation.}}

\author{Nils K\"urbis\orcidID{0000-0002-3651-5458}}
\date{}

\authorrunning{Nils K\"urbis}

\institute{Dept. of Logic and Methodology of Science, University of \L \'od\'z, Poland \and
Dept. of Philosophie I, University of Bochum, Germany
\email{nils.kurbis@filhist.uni.lodz.pl}\\
\url{https://www.nilskurbis.weebly.com}
}

%Acknowledgements should generally be placed in an unnumbered subsection at the end of the paper. If you still need to refer to a support or funding program in a note to the title, you can use the\thanks macro inside the title:

\maketitle

\begin{abstract}
This paper presents a sequent calculus and a dual domain semantics for a theory of definite descriptions in which these expressions are formalised in the context of complete sentences by a binary quantifier $I$. $I$ forms a formula from two formulas. $Ix[F, G]$ means `The $F$ is $G$'. This approach has the advantage of incorporating scope distinctions directly into the notation. Cut elimination is proved for a system of classical positive free logic with $I$ and it is shown to be sound and complete for the semantics. The system has a number of novel features and is briefly compared to the usual approach of formalising `the $F$' by a term forming operator. It does not coincide with Hintikka's and Lambert's preferred theories, but the divergence is well-motivated and attractive.
\end{abstract}

\section{Introduction}
A definite description is an expression of the form `the $F$'. Accordingly, the most popular formalisations of the theory of definite descriptions treat them as term forming operators: the operator $\iota$ binds a variable and turns an open formula into a singular term $\iota xF$. This treatment of definite descriptions goes back to Whitehead and Russell \cite{PM1}.\footnote{\label{fregedefarticle}Frege's treatment of the function that is a `substitute for the definite article' is different. Frege's operator $\backslash$ applies to names of objects, not to (simple or complex) predicates or function symbols. Typically these names refer to the extensions of concepts, but this is not necessary. $\backslash\xi$ returns the unique object that falls under a concept, if $\xi$ is a name of the extension of a concept under which a unique object falls, and its argument in all other cases. See \cite[\S 11]{fregegrundgesetzeI}.} Whitehead and Russell, however, did not consider definite descriptions to be genuine singular terms: they only have meaning in the context of complete sentences in which they occur and disappear upon analysis: `The $F$ is $G$' is logically equivalent to `There is one and only one $F$ and it is $G$'. Following the work of Hinitkka \cite{hintikkatowardsdefdesc} and Lambert \cite{lambertnotesEII}, many logicians prefer to formalise definite descriptions in a fashion where they are not straightforwardly eliminable. In such systems, $\iota$ is governed by what has come to be called \emph{Lambert's Law}: 

\lbp{LL}{$LL$}{$\forall y(\iota xFx=y \leftrightarrow \forall x(Fx\leftrightarrow x=y))$}

\noindent The preferred logic of many free logicians is \emph{positive free logic}, where formulas containing names that do not refer (to objects considered amongst those that exist) may be true. Then `The $F$ is $G$' is no longer equivalent to `There is one and only one $F$ and it is $G$'. In \emph{negative free logic}, all atomic formulas containing non-denoting terms are false, and the Russellian analysis is again appropriate. 

There is agreement amongst free logicians that \rf{LL} formalises the minimal theory of definite descriptions. Lambert himself prefers a stronger theory \cite{lambertnotesEIII} that in addition has the axiom:\footnote{This axiom bears some resemblance to Frege's Basic Law VI, the sole axiom for his operator $\backslash$, which is $a=\backslash\acute{\varepsilon}(a=\varepsilon)$ \cite[\S 18]{fregegrundgesetzeI}. But see footnote \ref{fregedefarticle}.}

\lbp{FL}{$FL$}{$t=\iota x (x=t)$} 

\noindent There are a number of other axioms that have been considered, but these two will be the focus of the present investigation.\footnote{For a survey of various theories and their axioms, see \cite{bencivengahandbook,fraassenthexxlambert,lambertfoundations,morschersimonsfreelogic}.} The proof theory of the theory of definite descriptions has received close study from the hands of Andrzej Indrzejczak.\footnote{An earlier approach is by Czermak \cite{czermakdefdescr}. Gratzl provides a cut free proof system for Russell's theory of definite descriptions, including his method for marking scope \cite{gratzldefdescr}.} In a series of papers, Indrzejczak has investigated various formalisations of theories of definite descriptions and provided cut free sequent calculi for them \cite{andrzejmodaldescription,andrzejfregean,andrzejexistencedefinedness,andrzejfreedefdescr,andrzejrussellian}. A cut free system of positive free logic of his will form the background to the present paper. It is presented in the next section. 

Whitehead and Russell also note the need for marking scope distinctions to formalise the difference between `The $F$ is not $G$' and `It is not the case that the $F$ is $G$'. Free definite description theory in general ignores scope: the thought is that free logic says only very little about definite descriptions when they do not refer, and in case they do refer, scope distinctions no longer matter, as already pointed out by Whitehead and Russell. 

Scope distinctions are, however, worth considering. The present paper proposes a proof-system and a semantics for a theory of definite descriptions in which scope distinctions are incorporated directly into the symbolism. `The $F$ is $G$' is formalised by a binary quantifier that takes two formulas and forms a formula $Ix[F, G]$ out of them. The notation is taken from Dummett \cite[p.162]{dummettfregelanguage}. It is also found in the work of Neale \cite{nealedescriptions} and Bostock \cite[Sec. 8.4]{bostockintermediate}. The external negation `It is not the case that the $F$ is $G$' is formalised by $\neg Ix[F, G]$, the internal negation `The $F$ is not $G$' by $Ix[F, \neg G]$. Natural deduction proof-systems for this approach have been investigated by the present author in the context of intuitionist non-free as well as negative and positive free logic \cite{kurbisiotaI,kurbisiotaII,kurbisiotaIII}. Rules suitable for a sequent calculus for classical positive free logic were formulated in \cite{kurbisiotasequentI}.\footnote{This paper also briefly considers rules for classical non-free and negative free logic.} The latter system and its intuitionist counterpart were devised with the intention to stay close to the systems of Hintikka and Lambert. The results are rather complicated: $I$ is governed by six rules, one right or introduction rule and five left or elimination rules. Despite their complexities, the systems remain proof-theoretically satisfactory as cut elimination and normalisation theorems hold for them. The present paper severs the ties to Hintikka and Lambert and considers alternative rules for $I$ within classical positive free logic. The account proposed here is rather simpler than the previous ones: $I$ is governed by one right rule, the same as before, but only two left rules. The result is a rather different formal theory from the perspective of the validities provable from the rules and compared to Hintikka's and Lambert's: the rules enforce the uniqueness of $F$, if $Ix[F, G]$ is true, but not its existence. The novelty of the present paper lies in the addition of these new rules for $I$ to classical positive free logic,\footnote{The rules are, in fact, those given for non-free classical logic at the end of \cite{kurbisiotasequentI}: it is a noteworthy result that, whereas in the context of this logic these rules are redundant and $Ix[F, G]$ definable in Russellian fashion as %equivalent to 
$\exists x(F\land \forall y(F_y^x\rightarrow x=y)\land G)$, added to classical positive free logic, the outcome is a theory of considerable logical and philosophical interest.} the ensuing alternative theory of definite descriptions, and the provision of a sound and complete dual domain semantics for it. 

The plan of this paper is as follows. The next section expounds Indrzejczak's sequent calculus formulation of classical positive free logic extended by rules governing the binary quantifier $I$. Section 3 discusses consequences of the theory and compares it to Hintikka's and Lambert's. Due to the absence of scope distinctions in axiomatisations of $\iota$ based on \rf{LL}, a direct comparison between the system proposed here and standard formalisations of definite descriptions is not very illuminating: $G(\iota xF)$ has no direct and natural correspondent, as $\neg G(\iota xF)$ corresponds to two formulas, the internal and the external negation of $Ix[F, G]$. Nonetheless, it is worth examining how the binary fares with respect to analogues of\rf{LL} and \rf{FL}, when $\iota xA=y$ is rendered as a binary quantification $Ix[A, x=y]$. The latter formalises `The $A$ is identical to $y$', or `The $A$ is $y$' for short, which is exactly the reading one may give of $\iota xA=y$. To anticipate, while an analogue of \rf{FL} is derivable in the system proposes here, only half of an analogue of \rf{LL} is. Section 4 proves that cut is still eliminable from the extended system. Section 5 gives a formal semantics for classical positive free logic extended by $I$. Section 6 proves the soundness and completeness of the system. Some details of the completeness proof are relegated to the Appendix. Section 7 gives rules tableaux proof system.

\section{A Deductive Calculus for Classical Positive Free Logic with a Binary Quantifier}
Indrzejczak has provided a formalisation of classical positive free logic \textbf{CPF} in sequent calculus with desirable proof-theoretic properties: cut is eliminable from the system \cite{andrzejcutfreefreelogic}. The definition of the language is standard. I will only consider $\rightarrow$, $\neg$, $\forall$ and a distinguished predicate symbol $\exists!$, the existence predicate, as primitives.\footnote{It would be possible to define $\exists !t$ as $\exists x \ x=t$, where $\exists$ may in turn be defined in terms of $\forall$ and $\neg$. However, treating it as primitive is formally and philosophically preferable: formally, it lends itself more easily to cut elimination, and philosophically, it permits to take existence as conceptually basic, with the quantifiers explained in terms of it: the attempted definition of $\exists!$ is arguably circular, as the rules of inference governing $\forall$, which explain its meaning, appeal to $\exists!$. The semantic clause for $\forall$, too, implicitly appeals to the concept of existence, as it ranges only over objects in the domain of the model which are considered to exist, that is, those of which $\exists!$ is true.} $\land$, $\lor$, $\exists$ are defined as usual. Free variables are distinguished from bound ones by the use of parameters $a, b, c\ldots$ for the former and $x, y, z \ldots$ for the latter. $t_1, t_2, t_3\ldots$ range over the terms of the language, which are the parameters, constants, and complex terms formed from them and function symbols. For brevity I will write $F$ or $A$ instead of $F(x)$ or $A(x)$ etc., except in the case of the existence predicate, where I'll write $\exists !x$ etc.. $A_t^x$ is the result of substituting $t$ for $x$ in $A$, where it is assumed that no variable free in $t$ becomes bound in $A_t^x$, i.e. that $t$ is free for $x$ in $A$. $\Gamma, \Delta$ denote finite multisets of formulas. The rules of \textbf{CPF} are as follows: 

\def\fCenter{\ \Rightarrow\ }

\begin{longtable}{l l}
(Ax) \ \AxiomC{$A\fCenter A$}
\DisplayProof & 
\AxiomC{$\Gamma\Rightarrow\Theta, A$}
\AxiomC{$A, \Delta\Rightarrow \Lambda$}
\LeftLabel{Cut \ }
\BinaryInfC{$\Gamma, \Delta\Rightarrow\Theta, \Lambda$}
\DisplayProof\\
 \\
\Axiom$\Gamma\fCenter \Delta$
\LeftLabel{(LW)\ }
\UnaryInf$A, \Gamma\fCenter \Delta$
\DisplayProof  &
\Axiom$\Gamma\fCenter \Delta$
\LeftLabel{(RW) \ }
\UnaryInf$\Gamma\fCenter \Delta, A$
\DisplayProof\\
\\
\Axiom$A, A, \Gamma\fCenter \Delta$
\LeftLabel{(LC) \ }
\UnaryInf$A, \Gamma\fCenter \Delta$
\DisplayProof & 
\Axiom$\Gamma\fCenter \Delta, A, A$
\LeftLabel{(RC) \ }
\UnaryInf$\Gamma\fCenter\Delta, A$
\DisplayProof\\
\\
\Axiom$\Gamma\fCenter \Delta, A$
\LeftLabel{$(L\neg)$ \ }
\UnaryInf$\neg A, \Gamma\fCenter \Delta$
\DisplayProof &
\Axiom$A, \Gamma\fCenter \Delta$
\LeftLabel{$(R\neg)$ \ }
\UnaryInf$\Gamma\fCenter \Delta, \neg A$
\DisplayProof\\
\\
\AxiomC{$\Gamma\Rightarrow \Delta, A$} 
\AxiomC{$B, \Gamma\Rightarrow \Delta$}
\LeftLabel{$(L\!\rightarrow)$ \ }
\BinaryInfC{$A\rightarrow B, \Gamma\Rightarrow \Delta$}
\DisplayProof & 
\Axiom$A, \Gamma\fCenter \Delta, B$
\LeftLabel{$(R\!\rightarrow)$ \ }
\UnaryInf$\Gamma\fCenter \Delta, A\rightarrow B$
\DisplayProof\\
\\
\Axiom$A_t^x, \Gamma\fCenter \Delta$
\LeftLabel{$(L\forall)$ \ }
\UnaryInf$\exists !t, \forall xA, \Gamma \fCenter \Delta$
\DisplayProof & 
\Axiom$\exists ! a, \Gamma \fCenter \Delta, A_a^x$
\LeftLabel{$(R\forall)$ \ }
\UnaryInf$\Gamma\fCenter \Delta, \forall xA$
\DisplayProof\\
\\
\Axiom$A_{t_2}^x, \Gamma\fCenter\Delta$
\LeftLabel{$(=I)$ \ }
\UnaryInf$t_1=t_2, A_{t_1}^x, \Gamma\fCenter\Delta$
\DisplayProof & 
\Axiom$t=t, \Gamma\fCenter\Delta$
\LeftLabel{$(=E)$ \ }
\UnaryInf$\Gamma\fCenter\Delta$
\DisplayProof
\end{longtable}

\noindent where in $(R\forall)$, $a$ does not occur in the conclusion, and in $(L\forall)$, $t$ is substitutable for $x$ in $A$. In $(=I)$, $A$ is atomic. The general case follows by induction. 

To these we add rules for the binary quantifier $I$:\bigskip

\AxiomC{$\Gamma\Rightarrow \Delta, A_t^x$}
\AxiomC{$\Gamma\Rightarrow \Delta, B_t^x$}
\AxiomC{$A_a^x, \Gamma \Rightarrow \Delta, a=t$}
\LeftLabel{$(RI)$ \ }
\TrinaryInfC{$\Gamma \Rightarrow \Delta, Ix[A, B]$}
\DisplayProof

\bigskip

\Axiom$A_a^x, B_a^x, \Gamma \fCenter \Delta$
\LeftLabel{$(LI^1)$ \ }
\UnaryInf$Ix[A, B], \Gamma\fCenter\Delta$
\DisplayProof

\bigskip

\AxiomC{$\Gamma\Rightarrow \Delta, A_{t_1}^x$}
\AxiomC{$\Gamma\Rightarrow \Delta, A_{t_2}^x$}
\AxiomC{$\Gamma\Rightarrow \Delta, C_{t_2}^x$}
\LeftLabel{$(LI^2)$ \ }
\TrinaryInfC{$Ix[A, B], \Gamma \Rightarrow \Delta, C_{t_1}^x$}
\DisplayProof

\bigskip

\noindent where in $(RI)$ and $(LI^1)$, $a$ does not occur in the conclusion, and in $(LI^2)$ $C$ is an atomic formula. The general case follows by induction.

Vacuous quantification with $I$ is allowed. If $x$ is not free in $A$, then the truth of $Ix[A, B]$ requires or imposes a restriction on the domain: if there is only one object (existing or not), then, if $A$ is true and $B$ is true (of the object in the domain, if $x$ is free in $B$), then $Ix[A, B]$ is true; and if $Ix[A, B]$ is true, then, if $A$ is true, then there is only one object in the domain and $B$ is true (of it, if $x$ is free in $B$). If $x$ is not free in $B$, then $Ix[A, B]$ is true if and only if a unique object (existing or not) is $A$ and $B$ is true. 

Call the resulting system $\mathbf{CPF}^I$. Deductions are defined as usual, as certain trees with axioms at the top-nodes or leaves and the conclusion at the bottom-node or root. If a sequent $\Gamma\Rightarrow \Delta$ is deducible in $\mathbf{CPF}^I$, we write $\vdash \Gamma\Rightarrow \Delta$.

\section{Consequences of the Formalisation}
Call two formulas $\iota xA=y$ and $Ix[A, x=y]$ \emph{analogues} of each other. They both formalise the same sentence `The $A$ is identical to $y$'. Similarly for $B(\iota xA)$ and $Ix[A, B]$, where we restrict $B$ to atomic formulas to avoid complications regarding scope. Let $\mathbf{CPF}^\iota$ be $\mathbf{CPF}$ plus \rf{LL} and \rf{FL}. Analogues provide a convenient means for comparisons between $\mathbf{CPF}^I$ and $\mathbf{CPF}^\iota$. 

First we state the obvious. The Law of Identity $\Rightarrow t=t$ and Leibniz' Law $t_1=t_2, A_{t_2}^x \Rightarrow A_{t_1}^x$ are derivable in \textbf{CPF}:\bigskip

\def\fCenter{\ \Rightarrow\ }
\Axiom$t=t \fCenter t=t$
\UnaryInf$ \fCenter t=t$
\DisplayProof\qquad \qquad
\def\fCenter{\ \Rightarrow\ }
\Axiom$A_{t_1}^x \fCenter A_{t_1}^x$
\UnaryInf$t_2=t_1, A_{t_2}^x \fCenter A_{t_1}^x$
\UnaryInf$t_1=t_2, t_1=t_1, A_{t_2}^x \fCenter A_{t_1}^x$
\UnaryInf$t_1=t_2, A_{t_2}^x \fCenter A_{t_1}^x$
\DisplayProof\bigskip

\noindent $t_1=t_2, A_{t_1}^x \Rightarrow A_{t_2}^x$ of course also holds, as established by step two of the left deduction through interchanging $t_1$ and $t_2$. 

Leibniz' Law is no longer applicable to definite descriptions in the present framework, as definite descriptions are not analysed as singular terms but only in the context of complete sentences in which they occur. We can, however, mimic its use, as we can derive the sequents $Ix[A, x=t], B_t^x\Rightarrow Ix[A, B]$, $Ix[A, x=t], Ix[A, B],\Rightarrow B_t^x$ and $Ix[A, Iy[B, x=y]], Ix[A, C]\Rightarrow Ix[B, C]$. Using analogues, these correspond to instances of Leibniz' Law: $\iota xA=t, B_t^x\Rightarrow B(\iota xA)$, $\iota xA=t, B(\iota xA),\Rightarrow B_t^x$ and $\iota xA=\iota yB, C(\iota xA)\Rightarrow C(\iota yB)$. We'll prove the first for purposes of illustration. Double lines indicate applications of structural rules, in particular those needed to make the contexts of the rules identical by Thinning. Let $\Pi$ be the following deduction in $\mathbf{CPF}^I$: 

\begin{prooftree}
\def\fCenter{\ \Rightarrow\ }
\Axiom$A_b^x\fCenter A_b^x$
\Axiom$a=t, A_a^x\fCenter A_t^x$
\Axiom$\fCenter t=t$
\doubleLine
\TrinaryInf$Ix[A, x=t], A_b^x, a=t, A_a^x\fCenter b=t$
\UnaryInf$Ix[A, x=t], Ix[A, x=t], A_b^x\fCenter b=t$
\UnaryInf$Ix[A, x=t], A_b^x\fCenter b=t$
\end{prooftree} 

\noindent Then the following establishes the analogue of our instance of Leibniz' Law: 

\begin{prooftree} 
\def\fCenter{\ \Rightarrow\ }
\Axiom$a=t, A_a^x\fCenter A_t^x$
\Axiom$B_t^x\fCenter B_t^x$
\AxiomC{$\Pi$}
\doubleLine
\TrinaryInf$a=t, A_a^x, B_t^x, Ix[A, x=t]\fCenter Ix[A, B]$
\UnaryInf$Ix[A, x=t], B_t^x, Ix[A, x=t]\fCenter Ix[A, B]$
\UnaryInf$Ix[A, x=t], B_t^x\fCenter Ix[A, B]$
\end{prooftree} 

\noindent To assess whether \rf{LL} is provable, it is useful to have rules for the biconditional: 

\begin{center} 
\AxiomC{$\Gamma\Rightarrow \Delta, A, B$} 
\AxiomC{$A, B, \Gamma\Rightarrow \Delta$}
\LeftLabel{$(L\leftrightarrow)$ \ }
\BinaryInfC{$A\leftrightarrow B, \Gamma\Rightarrow \Delta$}
\DisplayProof\bigskip

\AxiomC{$A, \Gamma\Rightarrow \Delta, B$}
\AxiomC{$B, \Gamma\Rightarrow \Delta, A$}
\LeftLabel{$(R\leftrightarrow)$ \ }
\BinaryInfC{$\Gamma\Rightarrow \Delta, A\leftrightarrow B$}
\DisplayProof
\end{center} 

\noindent These are derivable from the rules for $\rightarrow$ given the usual definition of $\leftrightarrow$. 

Next we derive one half of an analogue of \rf{LL} in $\mathbf{CPF}^I$: 

{\small
\begin{prooftree} 
\def\fCenter{\ \Rightarrow\ }
\Axiom$A_a^x, a=b\fCenter A_b^x$ 
\Axiom$A_c^x\fCenter A_c^x$ 
\Axiom$\fCenter c=c$
\doubleLine
\TrinaryInf$Ix[A, x=b], A_a^x, a=b, A_c^x\fCenter c=b$
\Axiom$A_a^x, a=b, c=b\fCenter A_c^x$ 
\doubleLine
\BinaryInf$Ix[A, x=b], A_a^x, a=b \fCenter A_c^x\leftrightarrow c=b$ 
\UnaryInf$\exists ! c, Ix[A, x=b], A_a^x, a=b \fCenter A_c^x\leftrightarrow c=b$
\UnaryInf$Ix[A, x=b], A_a^x, a=b \fCenter \forall x(A\leftrightarrow x=b)$
\UnaryInf$Ix[A, x=b], Ix[A, x=b] \fCenter \forall x(A\leftrightarrow x=b)$
\UnaryInf$Ix[A, x=b] \fCenter \forall x(A\leftrightarrow x=b)$
\UnaryInf$\fCenter Ix[A, x=b]  \rightarrow\forall x(A\leftrightarrow x=b)$
\UnaryInf$\exists ! b\fCenter Ix[A, x=b]  \rightarrow \forall x(A\leftrightarrow x=b)$
\UnaryInf$\fCenter \forall y(Ix[A, x=y]  \rightarrow\forall x(A\leftrightarrow x=y))$
\end{prooftree}
}

\noindent The left and rightmost leaves are derivable by Leibniz' Law. 

The other half of \rf{LL} is not derivable in $\mathbf{CPF}^I$. Intuitively, there being a unique \emph{existing} $A$ is not sufficient for $Ix[A, B]$, as there may also be \emph{non-existing} $A$s in addition. It is straightforward to give a countermodel with the semantics of Section 5. 

$\Rightarrow Ix[x=t, x=t]$ follows by twice the Law of Identity and one application of $(RI)$, where both $A$ and $B$ are $x=t$: 

\begin{prooftree}
\def\fCenter{\ \Rightarrow\ }
\Axiom$ \fCenter t=t$
\Axiom$ \fCenter t=t$
\Axiom$a=t \fCenter a=t$
\TrinaryInf$\fCenter  Ix[x=t, x=t]$
\end{prooftree} 

\noindent Thus the analogue of \rf{FL} is derivable in $\mathbf{CPF}^I$. This is worth noting: Lambert calls \rf{FL} `an important theorem in traditional description theory' \cite[58]{lambertnotesEIII}, and, not being derivable in the minimal theory, is forced to add it as a further axiom. 

The present theory of definite descriptions is thus not comparable to Hintikka's and Lambert's minimal theory: it contains only one half of the analogue \rf{LL}, but also the analogue of \rf{FL}. The first respect provides a sense in which the present theory is weaker than Lambert's preferred theory, the second one in which it is stronger, because the rules for $I$ and $=$ yield the analogue of \rf{FL} immediately, while in Lambert's theory, \rf{FL} needs to be added as an extra axiom governing the definite description operator $\iota$. The novelty of the present theory is shown by these features. In particular, the failure of the right to left half of \rf{LL} is, arguably and \emph{pace} Hintikka and Lambert, desirable, for the reason stated. 

The theory does not allow the derivation of the analogue of $\iota xF=\iota xF$, $Ix[F, Iy[F, x=y]]$. This is a tolerable loss. As Russell is not identical to Whitehead, it is not difficult to accept that `The author of \emph{Principia Mathematica} $=$ the author of \emph{Principia Mathematica}' is not logically true. Reasons normally given for accepting $\iota xF=\iota xF$ is that it is an instance of the Law of Identity. These reasons, however, are not conclusive, as the example shows. $Ix[F, Iy[F, x=y]]$ is not an instance of the Law of Identity, and hence accepting that law does not force us to accept it. If more than two objects satisfy $F$, then it is false. 

Its differences to Hintikka's and Lambert's theory of definite descriptions are advantages of the present proposal. It allows us to reject the claim that the author of \emph{Principia Mathematica} is identical to the author of \emph{Principia Mathematica} and to declare `The author of \emph{Principia Mathematica} smokes a pipe' to be false. If there is more than one $A$, existing or not, then $Ix[A, B]$ is false, whatever $B$ may be: an identity, a predicate letter, a complex formula. The present theory provides principled reasons for declaring certain sentences containing definite descriptions to be false on which Hintikka and Lambert prefer to remain silent and for not having to accept some sentences they pronounce as logically true on grounds which one may well want to reject.

\section{Cut Elimination for $\mathbf{CPF}^I$}
We'll continue Indrzejczak's proof of cut elimination for $\mathbf{CPF}$ by adding the cases covering $I$. Let $d(A)$ be the degree of the formula $A$, that is the number of connectives occurring in it. $\exists ! t$ is atomic, that is of degree $0$. $d(\mathcal{D})$ is the degree of the highest degree of any cut formula in deduction $\mathcal{D}$. $A^k$ denotes $k$ occurrences of $A$, $\Gamma^k$ $k$ occurrences of the formulas in $\Gamma$. The height of a deduction is the largest number of rules applied above the conclusion, that is the number of nodes of a longest branch in the deduction. The proof appeals to the Substitution Lemma: 

\begin{lemma}
If $\vdash_k \Gamma\Rightarrow \Delta$, then $\vdash_k \Gamma_t^a\Rightarrow \Delta_t^a$. 
\end{lemma} 

\noindent Its proof goes through as usual. Consequently, we can always rewrite deductions so that each application of $(R\forall)$, $(RI)$ and $(LI^1)$ has its own parameter that occurs nowhere else in the proof. In the following, it will be tacitly assumed that deductions have been treated accordingly. 

\begin{lemma}[Right Reduction]
If $\mathcal{D}_1\vdash \Theta\Rightarrow \Lambda, A$, where $A$ is principal, and $\mathcal{D}_2\vdash A^k, \Gamma\Rightarrow\Delta$ have degrees $d(\mathcal{D}_1), d(\mathcal{D}_2)<d(A)$, then there is a proof $\mathcal{D}\vdash \Theta^k, \Gamma\Rightarrow \Lambda^k, \Delta$ with $d(\mathcal{D})<d(A)$. 
\end{lemma}

\noindent \emph{Proof.} By induction over the height of $\mathcal{D}_2$. The basis is trivial: if $d(\mathcal{D}_2)=1$, then $A^k, \Gamma\Rightarrow\Delta$ is an axiom and hence $k=1$, $\Gamma$ is empty, and $\Delta$ consists of only one $A$; we need to show $\Theta\Rightarrow \Lambda, A$, but that is already proved by $\mathcal{D}_1$. 

For the induction step, we consider the rules for $I$:\bigskip

\noindent (I) The last step of $\mathcal{D}_2$ is by $(RI)$. Then the occurrences $A^k$ in the conclusion of $\mathcal{D}_2$ are parametric and occur in all three premises of $(RI)$: apply the induction hypothesis to them and apply $(RI)$ afterwards. The result is the desired proof $\mathcal{D}$.\bigskip

\noindent (II) The last step of $\mathcal{D}_2$ is by $(LI^1)$. There are two options: 

\noindent (II.a) If the principal formula $Ix[F, G]$ of $(LI^1)$ is not one of the $A^k$, then apply the induction hypothesis to the premises of $(LI^1)$ and then apply $(LI^1)$. 

\noindent (II.b) If the principal formula $Ix[F, G]$ of $(LI^1)$ is one of the $A^k$, then $\mathcal{D}_2$ ends with: 

\begin{center} 
\Axiom$F_a^x, G_a^x, Ix[F, G]^{k-1}, \Gamma\fCenter \Delta$
\UnaryInf$Ix[F, G]^k, \Gamma\fCenter\Delta$
\DisplayProof
\end{center}

\noindent By induction hypothesis there is a deduction of $ F_a^x, G_a^x, \Theta^{k-1}, \Gamma\Rightarrow \Lambda^{k-1}, \Delta$ with cut degree less than $d(A)$, and by the Substitution Lemma:\bigskip

(1) $F_t^x,  G_t^x, \Theta^{k-1}, \Gamma\Rightarrow \Lambda^{k-1}, \Delta$\bigskip

\noindent $A$, i.e. $Ix[F, G]$, is principal in $\mathcal{D}_1$, so it ends with an application of $(RI)$: 

\begin{center} 
\AxiomC{$\Theta\Rightarrow \Lambda, F_t^x$}
\AxiomC{$\Theta\Rightarrow \Lambda, G_t^x$}
\AxiomC{$F_a^x, \Theta \Rightarrow \Lambda, a=t$}
\TrinaryInfC{$\Theta \Rightarrow \Lambda, Ix[F, G]$}
\DisplayProof
\end{center} 

\noindent Apply two cuts with (1) and the first and second premise, and conclude by contraction $\Theta^k, \Gamma\Rightarrow \Lambda^k, \Delta$.

\bigskip

\noindent (III) The last step of $\mathcal{D}_2$ is by $(LI^2)$. In this case the succedent of the conclusion of $\mathcal{D}_2$ is $\Delta, C_{t_1}$, where $C_{t_1}$ is an atomic formula. There are two cases. 

\noindent (III.a) The principal formula $Ix[F, G]$ of $(LI^2)$ is not one of the $A^k$: apply the induction hypothesis to the premises of $(LI^2)$ and then apply the rule.

\noindent (III.b) The principal formula $Ix[F, G]$ of $(LI^2)$ is one of the $A^k$.  Then $\mathcal{D}_2$ ends with: 

\begin{center}
\AxiomC{$Ix[F, G]^{k-1}, \Gamma\Rightarrow \Delta, F_{t_1}^x$}
\AxiomC{$Ix[F, G]^{k-1}, \Gamma\Rightarrow \Delta, F_{t_2}^x$}
\AxiomC{$Ix[F, G]^{k-1}, \Gamma\Rightarrow \Delta, C_{t_2}^x$}
\TrinaryInfC{$Ix[F, G]^k, \Gamma \Rightarrow \Delta, C_{t_1}$}
\DisplayProof
\end{center} 

\noindent By induction hypothesis, we have:

\bigskip 

(1) $\Theta^{k-1}, \Gamma\Rightarrow \Lambda^{k-1}, \Delta, F_{t_1}^x$

(2) $\Theta^{k-1}, \Gamma\Rightarrow \Lambda^{k-1}, \Delta, F_{t_2}^x$

(3) $\Theta^{k-1}, \Gamma\Rightarrow \Lambda^{k-1}, \Delta, C_{t_2}^x$

\bigskip

\noindent $A$, i.e. $Ix[F, G]$, is principal in $\mathcal{D}_1$, so it ends with an application of $(RI)$:  

\begin{center} 
\AxiomC{$\Theta\Rightarrow \Lambda, F_t^x$}
\AxiomC{$\Theta\Rightarrow \Lambda, G_t^x$}
\AxiomC{$F_a^x, \Theta \Rightarrow \Lambda, a=t$}
\TrinaryInfC{$\Theta \Rightarrow \Lambda, Ix[F, G]$}
\DisplayProof
\end{center} 

\noindent The Substitution Lemma applied to the third premise gives \bigskip

(5) $F_{t_1}^x, \Theta \Rightarrow \Lambda, t_1=t$ 

(6) $F_{t_2}^x, \Theta \Rightarrow \Lambda, t_2=t$\bigskip

\noindent To show: $\vdash \Theta^k, \Gamma\Rightarrow\Lambda^k, \Delta, C_{t_1}$ with $d(\mathcal{D})<d(Ix[F, G])$. Leibniz' Law gives\bigskip

(7) $t_1=t, t_2=t\Rightarrow t_1=t_2$

(8) $C_{t_2}, t_1=t_2\Rightarrow C_{t_1}$\bigskip

\noindent Cuts with (1) and (5) and with (2) and (6) give $\Theta^k, \Gamma\Rightarrow \Lambda^k, \Delta, t_1=t$ and $\Theta^k, \Gamma\Rightarrow \Lambda^k, \Delta, t_2=t$, whence by Cut with (7) and contraction $\Theta^k, \Gamma\Rightarrow \Lambda^k, \Delta, t_1=t_2$, and from the latter by Cuts with (3) and (8) and contraction $\Theta^k, \Gamma\Rightarrow\Lambda^k, \Delta, C_{t_1}$. As $C_{t_2}$ in $(LI^2)$ is restricted to atomic formulas, the degree of the ensuing deduction is less than $d(A)$, i.e. $d(Ix[F, G])$, which was to be proved.\bigskip

\noindent This completes the proof of the Right Reduction Lemma.

\begin{lemma}[Left Reduction]
If $\mathcal{D}_1\vdash \Gamma\Rightarrow \Delta, A^k$ and $\mathcal{D}_2\vdash A, \Theta\Rightarrow \Lambda$ have degrees $d(\mathcal{D}_1), d(\mathcal{D}_2)<d(A)$, then there is a proof $\mathcal{D}\vdash \Gamma, \Theta^k\Rightarrow \Delta, \Lambda^k$ with $d(\mathcal{D})<d(A)$. 
\end{lemma}

\noindent \emph{Proof} by induction over the height of $\mathcal{D}_1$. The basis is trivial, as then $\mathcal{D}_1$ is an axiom, and $\Gamma$ consists of one occurrence of $A$ and $\Delta$ is empty. What needs to be shown is that $A, \Theta\Rightarrow \Lambda$, which is already given by $\mathcal{D}_2$.

For the induction step, we distinguish two cases, and again we continue Indrzejczak's proof by adding the new cases arising through the addition of $I$.\bigskip

\noindent (A) No $A^k$ in the succedent of the conclusion of $\mathcal{D}_1$ is principal. Then we apply the induction hypothesis to the premises of the final rule applied in $\mathcal{D}_1$ and apply the final rule once more.\bigskip 

\noindent (B) Some $A^k$ in the succedent of the conclusion of $\mathcal{D}_1$ is principal. Two options:\bigskip 

\noindent (I) The final rule applied in $\mathcal{D}_1$ is $(RI)$: 

{\small
\begin{center}
\AxiomC{$\Gamma\Rightarrow \Delta, Ix[F, G]^{k-1}, F_t^x$}
\AxiomC{$\Gamma\Rightarrow \Delta, Ix[F, G]^{k-1}, G_t^x$}
\AxiomC{$F_a^x, \Gamma \Rightarrow \Delta, Ix[F, G]^{k-1}, a=t$}
\TrinaryInfC{$\Gamma\Rightarrow \Delta, Ix[F, G]^k$}
\DisplayProof
\end{center}
}

\noindent By induction hypothesis, we have\bigskip

(1) $\Gamma, \Theta^{k-1} \Rightarrow \Delta, \Lambda^{k-1}, F_t^x$

(2) $\Gamma, \Theta^{k-1}\Rightarrow \Delta, \Lambda^{k-1}, G_t^x$

(3) $F_a^x, \Gamma, \Theta^{k-1}\Rightarrow \Delta, \Lambda^{k-1}, a=t$\bigskip

\noindent Apply $(RI)$ with (1) to (3) as premises to conclude\bigskip 

(4) $\Gamma, \Theta^{k-1} \Rightarrow \Delta, \Lambda^{k-1}, Ix[F, G]$\bigskip 

\noindent Here $Ix[F, G]$ is principal, so apply the Right Reduction Lemma to the deduction concluding (4) and $\mathcal{D}_2$ (where $k=1$) to conclude $\Gamma, \Theta^k \Rightarrow \Delta, \Lambda^k$. \bigskip

\noindent (II) The final rule applied in $\mathcal{D}_1$ is $(LI^2)$:\bigskip

\AxiomC{$\Gamma\Rightarrow \Delta, {C_{t_1}}^{k-1}, F_{t_1}^x$}
\AxiomC{$\Gamma\Rightarrow \Delta, {C_{t_1}}^{k-1}, F_{t_2}^x$}
\AxiomC{$\Gamma\Rightarrow \Delta, {C_{t_1}}^{k-1}, C_{t_2}^x$}
\TrinaryInfC{$Ix[F, G], \Gamma \Rightarrow \Delta, {C_{t_1}}^k$}
\DisplayProof

\bigskip

\noindent By induction hypothesis, we have\bigskip

(1) $\Gamma, \Theta^{k-1}\Rightarrow \Delta, \Lambda^{k-1}, F_{t_1}^x$

(2) $\Gamma, \Theta^{k-1}\Rightarrow \Delta, \Lambda^{k-1}, F_{t_2}^x$

(3) $\Gamma, \Theta^{k-1}\Rightarrow \Delta, \Lambda^{k-1}, C_{t_2}^x$\bigskip

\noindent Apply $(LI^2)$ with (1) to (3) as premises to conclude:\bigskip

(4) $Ix[F, G], \Gamma, \Theta^{k-1}\Rightarrow \Delta, \Lambda^{k-1}, C_{t_1}^x$\bigskip

\noindent Here $C_{t_1}^x$ is principal, so apply the Right Reduction Lemma to the deduction concluding (4) and $\mathcal{D}_2$ (where $k=1$) to conclude $Ix[F, G], \Gamma, \Theta^k \Rightarrow \Delta, \Lambda^k$.\bigskip

\noindent This completes the proof of the Left Reduction Lemma. 

\begin{theorem}[Cut Elimination]
For every deduction in $\mathbf{CPF}^I$, there is a deduction that is free of cuts. 
\end{theorem}

\noindent \emph{Proof}. The theorem follows from the Right and Left Reduction Lemmas by induction over the degree of the proof, with subsidiary inductions over the number of cut formulas of highest degree, as in Indrzejczak's paper.

\section{Semantics for $\mathbf{CPF}^I$}
For the purposes of providing a semantics for $\mathbf{CPF}^I$ it is convenient to modify the system slightly in the following way: free variables $x, y, z \ldots$ are allowed to occur in formulas, parameters are treated like constants, and constants may play the role of parameters if they occur parametrically in a deduction, that is, they fulfil the restrictions imposed in $(R\forall)$, $(RI)$ and $(LI^1)$. The restrictions for free variables in these rules are as for the parameters. Furthermore, for the purposes of this section, I take $\Rightarrow$ to have sets of sentences rather the multisets to its left and right. I'll write $\Gamma, A$ to abbreviate $\Gamma\cup\{A\}$, $A, B, C\in \Delta$ for $\{A, B, C\}\subseteq\Delta$. The resulting modified system is evidently equivalent to the original formulation. 

It is fairly obvious that the rules governing $I$ enforce the uniqueness of $A$, if it is the case that $Ix[A, B]$, but not its existence. Arguing informally, it is immediate from $(LI^2)$ that $Ix[A, B], A_a^x, A_b^x\Rightarrow a=b$, hence any $A$s are identical; and if $A_a^x$ is false, whatever $a$ might be, then $A_a^x\Rightarrow \bot$, so by $(LI^1)$, $Ix[A, B]\Rightarrow \bot$. But the rules do not permit us to determine whether the unique $A$ exists or not. Conversely, to derive $Ix[A, B]$, we need a unique $A$ that is $B$, but it is not required that it exists. Nonetheless, we will prove it rigorously by providing a sound and complete semantics for $\mathbf{CPF}^I$. I follow the popular proposal by \cite{cocchiarellaobjects}, \cite{leblancthomason} and \cite{meyerlambert}, where two domains are considered, an inner one and an outer one, the former the domain of existing objects, over which the universal quantifier ranges and of which $\exists !$ is true, and the latter the domain of `non-existent' objects. I shall take the inner domain to be a subset of the outer domain. 

The exposition of the formal semantics for $\mathbf{CPF}^I$ and the soundness and completeness proofs in the next section follow Enderton closely, with necessary adjustments to be suitable to free logic. Most of the following is well known and not new, but I'll be explicit about the details in order to demonstrate the semantics of $I$ explicitly and precisely. 

A \emph{structure} $\mathfrak{A}$ is a function from the expressions of the language $\mathcal{L}$ of $\mathbf{CPF}^I$ to elements, a (possibly empty) subset, the sets of n-tuples of and operations on a non-empty set $|\mathfrak{A}|$, called the \emph{domain of} $\mathfrak{A}$, such that:\bigskip

\noindent 1. $\mathfrak{A}$ assigns to the quantifier $\forall$ a (possibly empty) set $|\mathfrak{A}^\forall|\subseteq|\mathfrak{A}|$ called the \emph{inner domain} or the \emph{domain of quantification} of $\mathfrak{A}$.

\noindent 2. $\mathfrak{A}$ assigns to the predicate $\exists!$ the set $|\mathfrak{A}^\forall|$.

\noindent 3. $\mathfrak{A}$ assigns to each $n$-place predicate symbol $P$ an $n$-ary relation $P^\mathfrak{A}\subseteq |\mathfrak{A}|^n$.

\noindent 4. $\mathfrak{A}$ assigns to each constant symbol $c$ an element $c^\mathfrak{A}$ of $|\mathfrak{A}|$.

\noindent 5. $\mathfrak{A}$ assigns to each $n$-place function symbol $f$ an $n$-ary operation $f^\mathfrak{A}$ on $|\mathfrak{A}|$, i.e. $f^\mathfrak{A}\colon |\mathfrak{A}|^n\to|\mathfrak{A}|$.\bigskip

\noindent Next we define the notion of \emph{satisfaction} of a formula $B$ by a structure $\mathfrak{A}$. To handle free variables we employ a function $s\colon V\to |\mathfrak{A}|$ from the set of variables $V$ of $\mathcal{L}$ to the domain of the structure. Suppose $x$ occurs free in $B$. Informally, we say that $\mathfrak{A}$ satisfies $B$ with $s$, if and only if the object of the domain of $\mathfrak{A}$ that $s$ assigns to the variable $x$ satisfies $B$, that is, if $s(x)$ is in the set $\mathfrak{A}$ assigns to $B$. We express this in symbols by $\vDash_\mathfrak{A}A\ [s]$. $\nvDash_\mathfrak{A}A\ [s]$ means that $\mathfrak{A}$ does not satisfy $A$ with $s$. The formal definition of satisfaction is as follows. 

First, $s$ is extended by recursion it to a function $\overline{s}$ that assigns objects of $|\mathfrak{A}|$ to all terms of the language:\bigskip 

\noindent 1. For each variable $x$, $\overline{s}(x)=s(x)$ 

\noindent 2. For each constant symbol $c$, $\overline{s}(c)=c^\mathfrak{A}$.  

\noindent 3. For terms $t_1\ldots t_n$, $n$-place function symbols $f$, $\overline{s}(ft_1\ldots t_n)=f^\mathfrak{A}(\overline{s}(t_1)\ldots \overline{s}(t_n))$\bigskip 

\noindent Satisfaction is defined explicitly for the atomic formulas of $\mathcal{L}$:\bigskip

\noindent 1. $\vDash_\mathfrak{A} t_1 = t_2 \ [s]$ iff $\overline{s}(t_1)=\overline{s}(t_2)$.  

\noindent 2. $\vDash_\mathfrak{A} \exists ! t  \ [s]$ iff $\overline{s}(t)\in|\mathfrak{A}^\forall|$. 

\noindent 3. For $n$-place predicate parameters $P$, $\vDash_\mathfrak{A} Pt_1\ldots t_n \ [s]$ iff $\langle \overline{s}(t_1)\ldots\overline{s}(t_n)\rangle\in P^\mathfrak{A}$.\bigskip 

\noindent For the rest of the formulas, satisfaction is defined by recursion. Let $s(x|d)$ be like $s$, only that it assigns the element $d$ of $|\mathfrak{A}|$ to the variable $x$:\bigskip

\noindent 1. For atomic formulas, as above. 

\noindent 2. $\vDash_\mathfrak{A} \neg A \ [s]$ iff $\nvDash_\mathfrak{A} A \ [s]$. 

\noindent 3. $\vDash_\mathfrak{A} A \rightarrow B \ [s]$ iff either $\nvDash_\mathfrak{A} A \ [s]$ or $\vDash_\mathfrak{A} B\ [s]$.  

\noindent 4. $\vDash_\mathfrak{A} \forall x A \ [s]$ iff for every $d\in|\mathfrak{A}^\forall|$, $\vDash_\mathfrak{A} A \ [s(x|d)]$. 
\bigskip

\noindent This gives a semantics for $\mathbf{CPF}$. For  $\mathbf{CPF}^I$, we add a clause for $I$:\bigskip 

\noindent 5. $\vDash_\mathfrak{A} Ix [A, B] \ [s]$ iff there is $d\in|\mathfrak{A}|$ such that: $\vDash_\mathfrak{A} A \ [s(x|d)]$, there is no other $e\in|\mathfrak{A}|$ such that $\vDash_\mathfrak{A} A \ [s(x|e)]$, and $\vDash_\mathfrak{A} B \ [s(x|d)]$.\bigskip

\noindent In other words, $\vDash_\mathfrak{A} Ix [F, G] \ [s]$ iff there is exactly one element in the domain of $\mathfrak{A}$ such that $\mathfrak{A}$ satisfies $A$ with $s$ modified to assign that element to $x$, and $\mathfrak{A}$ satisfies $B$ with the same modified $s$. 

We could define notions of validity, truth and falsity applicable to formulas, if we like, but won't need them in the following. A formula $A$ is \emph{valid} iff for every $\mathfrak{A}$ and every $s\colon V\to |\mathfrak{A}|$, $\vDash_\mathfrak{A} A \ [s]$. Call a formula with no free variables a sentence. A structure $\mathfrak{A}$ either satisfies a sentence $\sigma$ with every function $s\colon V\to |\mathfrak{A}|$ or with none. If the former, $\sigma$ is \emph{true} in $\mathfrak{A}$, if the latter, $\sigma$ is \emph{false} in $\mathfrak{A}$. If the former, we may write $\vDash_\mathfrak{A} \sigma$ and say that $\mathfrak{A}$ is a model of $\sigma$. 

More important are notions applicable to the sequents of the deductive system of $\mathbf{CPF}^I$. A sequent $\Gamma\Rightarrow \Delta$ is satisfied by a structure $\mathfrak{A}$ with a function $s\colon V\to |\mathfrak{A}|$ if and only if, if for all $A\in \Gamma$, $\vDash_\mathfrak{A} A \ [s]$, then for some $C\in\Delta$, $\vDash_\mathfrak{A} C \ [s]$. We symbolise this by $\vDash_\mathfrak{A} \Gamma\Rightarrow \Delta \ [s]$. A sequent $\Gamma\Rightarrow \Delta$ is \emph{valid} iff it is satisfied by every structure with every function $s\colon V\to |\mathfrak{A}|$. In this case we write $\vDash \Gamma\Rightarrow \Delta$. 

Sequents have finite sets to the left and right of $\Rightarrow$. We also need notions that apply to finite and infinite set. 

A set of formulas $\Gamma$ is \emph{satisfiable} iff there is some structure $\mathfrak{A}$ and some function $s\colon V\to |\mathfrak{A}|$ such that $\mathfrak{A}$ satisfies every member of $\Gamma$ with $s$. 

A set of formulas $\Gamma$ \emph{deductively implies} a formula $A$, iff for some finite $\Gamma_0\subseteq \Gamma$, $\vdash \Gamma\Rightarrow A$. If $\Gamma$ deductively implies $A$, we record this fact by $\Gamma\vdash A$. 

A set of formulas $\Gamma$ \emph{semantically implies} a formula $A$,  iff for every structure $\mathfrak{A}$ and every function $s\colon V\to |\mathfrak{A}|$ such that $\mathfrak{A}$ satisfies every member of $\Gamma$ with $s$, $\mathfrak{A}$ satisfies $A$ with $s$. If $\Gamma$ semantically implies $A$, we record this fact by $\Gamma\vDash A$.

\section{Soundness and Completeness}
I'll prove two pairs of soundness and completeness theorems: one pair shows that deducibility and validity of sequents coincide, and another that deductive and semantic implication coincide. 

A formula $A'$ is an \emph{alphabetic variant} of a formula $A$ if $A$ and $A'$ differ only in the choice of bound variables. 

\begin{lemma}[Existence of Alphabetic Variants]
For any formula $A$, term $t$ and variable $x$, there is a formula $A'$ such that $A\Rightarrow A'$ and $A'\Rightarrow A$ and $t$ is substitutable for $x$ in $A'$. 
\end{lemma}

\noindent \emph{Proof}. \emph{Mutatis mutandis} Enderton's proof goes through for $\mathbf{CPF}^I$, too \cite[126f]{endertonlogic}.\bigskip

\noindent Alphabetic variants are semantically equivalent: if $A$ and $A'$ are alphabetic variants, then $A\vDash A'$ and $A'\vDash A$.

\begin{lemma}[The Substitution Lemma.]
$\vDash_\mathfrak{A} A^x_t \ [s] \textrm{ iff } \vDash_\mathfrak{A} A \ [s(x|\overline{s}(t))]$, if $t$ is free for $x$ in $A$. 
\end{lemma} 

\noindent \emph{Proof.} See \cite[133f]{endertonlogic} and adjust. 

\begin{theorem}[Soundness for Sequents]\label{soundsequents}
If $\vdash \Gamma\Rightarrow \Delta$, then  $\vDash \Gamma\Rightarrow \Delta$. 
\end{theorem}

\noindent \emph{Proof.} Standard, by induction over the complexity of deductions and observing that the axioms are valid and all rules preserve validity. In the appendix, the soundness of the rules for the $\forall$ and $I$ is proved.

\begin{theorem}[Soundness for Sets]
If $\Gamma\vdash A$, then $\Gamma\vDash A$. 
\end{theorem}

\noindent \emph{Proof.} If $\Gamma\vdash A$, then for some finite $\Gamma_0\subseteq\Gamma$, $\vdash \Gamma_0\Rightarrow A$. So by Theorem \ref{soundsequents}, $\vDash \Gamma_0\Rightarrow A$. Suppose some structure $\mathfrak{A}$ satisfies all formulas of $\Gamma$ with a function $s\colon V\to |\mathfrak{A}|$. Then $\mathfrak{A}$ satisfies $\Gamma_o$ with $s$, hence, as $\vDash \Gamma_0\Rightarrow A$, $\mathfrak{A}$ satisfies $A$ with $s$, and so $\Gamma\vDash A$.\bigskip 

\noindent Some more definitions. Let $\bot$ represent an arbitrary contradiction. A set of formulas $\Gamma$ is \emph{inconsistent} iff $\Gamma\vdash \bot$. $\Gamma$ is \emph{consistent} iff it is not inconsistent. A set of formulas $\Gamma$ is \emph{maximal} iff for any formula $A$, either $A\in \Gamma$ or $\neg A\in \Gamma$. A set of formulas $\Gamma$ is \emph{deductively closed} iff, if $\Gamma\vdash A$, then $A\in\Gamma$. 

\begin{lemma}
Any maximally consistent set is deductively closed. 
\end{lemma}

\noindent\emph{Proof.} Suppose $\Gamma$ is maximal and $\Gamma\vdash A$ but $A\not\in\Gamma$. Then for some finite $\Gamma_0\subseteq\Gamma$, $\vdash \Gamma_0\Rightarrow A$. By maximality of $\Gamma$, $\neg A\in\Gamma$, hence for some finite $\Gamma_1\subseteq\Gamma$, $\vdash\Gamma_1\Rightarrow \neg A$. Hence $\vdash \Gamma_0, \Gamma_1\Rightarrow A\land \neg A$, and so $\Gamma\vdash\bot$. Contradiction. 

\begin{theorem}\label{maxconset}
Any consistent set  of formulas $\Delta$ can be extended to a maximally consistent set $\Delta^+$ such that: 

\noindent (a) for any formula $A$ and variable $x$, if $\neg\forall xA\in \Delta^+$, then for some constant $c$, $\exists !c\in\Delta^+$ and $A_c^x\not\in\Delta^+$; 

\noindent (b) for any formulas $A$ and $B$ and variable $x$, if $Ix[A, B]\in\Delta^+$, then for some constant $c$, $A_c^x, B_c^x\in\Delta^+$ and for all constants $d$, if $A_d^x\in\Delta^+$, then $d=c\in\Delta^+$. 

\noindent (c) for any formulas $A$ and $B$ and variable $x$, if $\neg Ix[A, B]\in\Delta^+$, then for all constants $c$, either $ A_c^x\not\in\Delta^+$, or for some constant $d$, $A_d^x\in\Delta^+$ and $d=c\not\in\Delta^+$, or $B_c^x\not\in\Delta^+$. 
\end{theorem}

\noindent \emph{Proof} is in the appendix. 

\begin{theorem}\label{satisfiability}
If $\Delta$ is a consistent set of formulas, then $\Delta$  is satisfiable. 
\end{theorem}

\noindent\emph{Proof} is in the appendix. 

\begin{theorem}[Completeness for Sequents]
If $\vDash \Gamma\Rightarrow \Delta$, then  $\vdash \Gamma\Rightarrow \Delta$. 
\end{theorem}

\noindent \emph{Proof.} Let $\neg \Delta$ be the negation of all formulas in $\Delta$. If $\vDash \Gamma\Rightarrow \Delta$, then $\Gamma, \neg \Delta$ is not satisfiable. Hence by Theorem \ref{satisfiability} it is inconsistent, and as they are both finite, $\vdash\Gamma, \neg\Delta  \Rightarrow \bot$. Hence by the properties of negation $\vdash \Gamma\Rightarrow \Delta$.

\begin{theorem}[Completeness for Sets]
If $\Gamma\vDash A$, then $\Gamma\vdash A$. 
\end{theorem}

\noindent \emph{Proof.} Suppose $\Gamma\vDash A$. Then $\Gamma, \neg A$ is not satisfiable, hence by Theorem \ref{satisfiability} it is inconsistent and $\Gamma, \neg A\vdash \bot$. So for some finite $\Sigma\subseteq \Gamma, \neg A$, $\Sigma\Rightarrow \bot$. If $\neg A\in \Sigma$, then by the deductive properties of negation, $\Sigma-\{\neg A\}\Rightarrow A$, and as $\Sigma-\{\neg A \}$ is certain to be a subset of $\Gamma$, $\Gamma\vdash A$. If $\neg A\not\in \Sigma$, then $\Sigma\Rightarrow A$ by the properties of negation, and again $\Gamma\vdash A$.

\section{Tableaux Rules}
In this section, we'll extend Priest's tableaux system for classical positive free logic \cite[Ch 13]{priestnonclassical} by rules for $I$. His rules give a system equivalent to $\mathbf{CPF}$: 

\begin{center} 
\Tree
[.$A\rightarrow B$ [.$\neg A$ ] [.$B$ ] ]
\qquad
\Tree
[.$\neg (A\rightarrow B)$ [.$\stackunder{A}{\neg B}$ ] ]
\qquad\qquad
\Tree 
[.$\neg\neg A$ [.$A$ ] ]
\qquad\qquad
\Tree
[.$\forall xA$ [.$\neg \exists !t$ ] [.$A_t^x$ ] ]
\qquad
\Tree
[.$\neg\forall xA$ [.$\stackunder{\exists !a}{\neg A_a^x}$ ] ]

\bigskip

\Tree
[. $\stackunder{A_{t_1}^x}{t_1=t_2}$ [.$\stackunder{\vert}{A_{t_2}^x}$ ] ]
\qquad
\Tree
[.  $\centerdot$ [.$t_3=t_3$ ] ]
\end{center}

\noindent where $t$ is any term on the branch (or a new one if there is none yet), $a$ is new to the branch and $t_3$ is any term.\bigskip

\noindent The binary quantifier $I$ has the following rules:

\begin{center}
\Tree
[.{${Ix[A, B]}$} [.$\stackunder{A_a^x}{B_a^x}$  [.$\neg A_t^x$ ] [.$a=t$ ] ] ] 
\qquad\qquad
\Tree
[.${\neg Ix[A, B]}$ [.$\neg A_t^x$ ] [.$\neg B_t^x$ ] [.$\stackunder{A_a^x}{\neg \  a=t}$ ] ]
\end{center} 

\noindent where $a$ is new to the branch and $t$ is any term on the branch (or a new one if there is none yet).

\section{Conclusion}
The theory of definite descriptions formulated here has some novel and attractive features. The proof-theory is simple and has desirable consequences. It differs from Hintikka's and Lambert's preferred theories in a well-motivated way. It lends itself to applications of formalisations in which scope distinctions are of importance. The distinction between internal and external negation has been mentioned in the introduction. Other, and particularly interesting, cases are found in modal discourse. There is a significant difference between `It is possible the that present King of France is bald' and `The present King of France is possibly bald'. In the present framework, the former is formalised by a formula such as $\Diamond Ix[Kx, Bx]$, the latter by $Ix[Kx, \Diamond Bx]$.  The importance of scope distinctions in the context of modal logic was first pointed out by Smullyan \cite{smullyanmoddesc}. His account was further developed by Hughes and Cresswell \cite[323ff]{hughescresswell}. Elaborate systems catering for definite descriptions in modal logic have been provided by  Fitting and Mendelsohn \cite{mendelsohnfitting} and Garson \cite{garsonmodallogic}. In both of the latter systems, an operator for predicate abstraction is used to mark scope, but it serves no further purpose. Future research will investigate the addition of the binary quantifier $I$ to quantified modal logic and compare the result to existing systems. In particular, as the present system incorporates scope distinctions directly into the formalism for representing definite descriptions, there is no need for additional means to mark scope. This promises economy and clarity in the formalism for representing definite descriptions where scope distinctions matter.

\subsection*{Acknowledgments}
I would like to thank Andrzej Indrzejczak for comments on this paper and discussions of the proof-theory of definite descriptions in general. Some of this material was presented at Heinrich Wansing's and Hitoshi Omori's Work in Progress Seminar at the University of Bochum, to whom many thanks are due for support and insightful comments. Last but not least I must thank the referees for \textsc{Tableaux 2021} for their thoughtful and considerate reports on this paper.

\section{Appendix. Proofs of Theorems \ref{soundsequents}, \ref{maxconset} and \ref{satisfiability}} 
\begin{manualtheorem}{\ref{soundsequents}}
If $\vdash \Gamma\Rightarrow \Delta$, then  $\vDash \Gamma\Rightarrow \Delta$. 
\end{manualtheorem}

\noindent \emph{Proof.} Here is the proof of soundness for the rules of $\forall$ and $I$.\bigskip 

\noindent $(L\forall)$. Suppose (1) $\vDash A_t^x, \Gamma\Rightarrow \Delta$, but (2) $\not\vDash_\mathfrak{A} \exists !t, \forall xA, \Gamma \Rightarrow \Delta$. Then by (2), there is a structure $\mathfrak{A}$ and a function $s\colon V\to |\mathfrak{A}|$ such that (3) $\vDash_\mathfrak{A} \exists ! t \ [s]$, (4) $\vDash_\mathfrak{A} \forall xA \ [s]$, for all $B\in\Gamma$, $\vDash_\mathfrak{A} B \ [s]$, and for all $C\in\Delta$, $\nvDash_\mathfrak{A} C \ [s]$. So by (1), (5) $\nvDash A_t^x \ [s]$. By (4), for all $d\in|\mathfrak{A}^\forall|$, $\vDash A \ [s(x|d)]$, and by (3), $\overline{s}(t)\in|\mathfrak{A}^\forall|$, so $\vDash A \ [s(x|\overline{s}(t))]$. The latter contradicts (5) by the Substitution Lemma and the conditions on $t$ in $(L\forall)$.\bigskip 

\noindent $(R\forall)$. Suppose (1) $\vDash \exists ! x, \Gamma\Rightarrow \Delta, A$, but (2) $\nvDash \Gamma\Rightarrow \Delta, \forall xA$. Then there is a structure $\mathfrak{A}$ and a function $s\colon V\to |\mathfrak{A}|$ such that for all $B\in\Gamma$, $\vDash_\mathfrak{A} B \ [s]$, for all $C\in\Delta$, $\nvDash_\mathfrak{A} C \ [s]$ and (3) $\nvDash_\mathfrak{A} \forall xA \ [s]$. By (3), for some $d\in|\mathfrak{A}^\forall|$, $\nvDash_\mathfrak{A} A \ [s(x|d)]$. As $x$ is not free in any formulas in $\Gamma$ or $\Delta$, $s$ and $s(x|d)$ agree for any of these formulas, and so for all $B\in\Gamma$, $\vDash_\mathfrak{A} B \ [s(x|d)]$ and for all $C\in\Delta$, $\nvDash_\mathfrak{A} C \ [s(x|d)]$ %could prove a lemma and refer to that one. 
Of course also, by (1), $\vDash_\mathfrak{A} \exists ! x, \Gamma\Rightarrow \Delta, A \ [s(x|d)]$, and so if $\vDash_\mathfrak{A} \exists ! x \ [s(x|d)]$, then (4) $\vDash_\mathfrak{A} A \ [s(x|d)]$. But $d\in|\mathfrak{A}^\forall|$, hence $\vDash_\mathfrak{A} \exists ! x \ [s(x|d)]$, and we reach a contradiction between (3) and (4).\bigskip

\noindent $(RI)$. Suppose (1) $\vDash \Gamma\Rightarrow \Delta, A_t^x$, (2) $\vDash \Gamma\Rightarrow \Delta, B_t^x$, (3) $\vDash A, \Gamma\Rightarrow \Delta, x=t$, but $\nvDash \Gamma\Rightarrow \Delta, Ix[A, B]$, where $x$ is not free in any formulas in $\Gamma$ and $\Delta$. Then by the last, there is a structure $\mathfrak{A}$ and a function $s\colon V\to |\mathfrak{A}|$ such that for all $C\in\Gamma$, $\vDash_\mathfrak{A} C \ [s]$, for all $D\in\Delta$, $\nvDash_\mathfrak{A} D \ [s]$ and (4) $\nvDash_\mathfrak{A} Ix[A, B]\ [s]$. So by (1), $\vDash_\mathfrak{A} A_t^x \ [s]$, by (2) $\vDash_\mathfrak{A} B_t^x \ [s]$. By (4) it is not the case that there is $d\in|\mathfrak{A}|$ such that: $\vDash_\mathfrak{A} A \ [s(x|d)]$, there is no other $e\in|\mathfrak{A}|$ such that $\vDash_\mathfrak{A} A \ [s(x|e)]$, and $\vDash_\mathfrak{A} B \ [s(x|d)]$, i.e. for every $d\in|\mathfrak{A}|$: either $\nvDash_\mathfrak{A} A \ [s(x|d)]$, or for some $e\in|\mathfrak{A}|$ other that $d$, $\vDash_\mathfrak{A} A \ [s(x|e)]$, or $\nvDash_\mathfrak{A} B \ [s(x|d)]$. Consider $\overline{s}(t)$. We have (5) $\nvDash_\mathfrak{A} A \ [s(x|\overline{s}(t))]$, or (6) for some $e\in|\mathfrak{A}|$ other that $\overline{s}(t)$, $\vDash_\mathfrak{A} A \ [s(x|e)]$, or (7) $\nvDash_\mathfrak{A} B \ [s(x|\overline{s}(t))]$. From (5) and (7) by the Substitution Lemma, $\nvDash_\mathfrak{A} A_t^x \ [s]$ and $\nvDash_\mathfrak{A} B_t^x \ [s]$, contradicting (1) and (2). This leaves (6). Consider the function that is just like $s$ but assigns $e$ to $x$. $x$ is not free in any formulas in $\Gamma$ and $\Delta$, so $s$ and $s(x|e)$ agree on all formulas in them, i.e. $C\in\Gamma$, $\vDash_\mathfrak{A} C \ [s(x|e)]$, for all $D\in\Delta$, $\nvDash_\mathfrak{A} D \ [s(x|e)]$. Hence by (3) if $\vDash_\mathfrak{A} A \ [s(x|e)]$, then  $\vDash_\mathfrak{A} x=t \ [s(x|e)]$. But $ e$ is different from $\overline{s}(t)$, hence $\nvDash_\mathfrak{A} x=t \ [s(x|e)]$, and so $\nvDash_\mathfrak{A} A \ [s(x|e)]$, contradicting (6). Overall contradiction. Hence $\vDash \Gamma\Rightarrow \Delta, Ix[A, B]$.\bigskip 

\noindent $(LI^1)$. Suppose (1) $\vDash A, B, \Gamma\Rightarrow \Delta$, but $\nvDash Ix[A, B], \Gamma\Rightarrow\Delta$, $x$ not free in $\Gamma, \Delta$. From the latter, there is a structure $\mathfrak{A}$ and a function $s\colon V\to |\mathfrak{A}|$ such that (2) $\vDash_\mathfrak{A} Ix[A, B]\ [s]$, for all $C\in\Gamma$, $\vDash_\mathfrak{A} C \ [s]$, for all $D\in\Delta$, $\nvDash_\mathfrak{A} D \ [s]$. As (2), there is exactly one $d\in|\mathfrak{A}|$ such that $\vDash_\mathfrak{A} A \ [s(x|d)]$ and for this $d$, (4) $\vDash_\mathfrak{A} B \ [s(x|d)]$. $x$ is not free in $\Gamma, \Delta$, so $s$ and $s(x|d)$ agree on all formulas in them. Hence from (1), either (3) $\nvDash_\mathfrak{A} A \ [s(x|d)]$ or (4) $\nvDash_\mathfrak{A} B \ [s(x|d)]$. Either way, contradiction. Hence $\vDash Ix[A, B], \Gamma\Rightarrow\Delta$.\bigskip

\noindent $(LI^2)$. Suppose (1) $\vDash \Gamma \Rightarrow \Delta, A_{t_1}$, (2) $\vDash \Gamma \Rightarrow \Delta, A_{t_2}$, (3) $\vDash \Gamma \Rightarrow \Delta, C_{t_2}$, but $\nvDash Ix[A, B], \Gamma \Rightarrow \Delta, C_{t_1}$. Then there is a structure $\mathfrak{A}$ and a function $s\colon V\to |\mathfrak{A}|$ such that (4) $\vDash_\mathfrak{A} Ix[A, B]\ [s]$, for all $C\in\Gamma$, $\vDash_\mathfrak{A} C \ [s]$, for all $D\in\Delta$, $\nvDash_\mathfrak{A} D \ [s]$ and $\nvDash_\mathfrak{A} C_{t_1} \ [s]$, from which by the Substitution Lemma  (5) $\nvDash_\mathfrak{A} C \ [s(x|\overline{s}(t_1))]$. So from (1) and (2), (6) $\vDash_\mathfrak{A} A_{t_1} \ [s]$ and (7) $\vDash_\mathfrak{A} A_{t_2} \ [s]$, and from (3), $\vDash_\mathfrak{A} C_{t_2} \ [s]$, from which by the Substitution Lemma (8) $\vDash_\mathfrak{A} C \ [s(x|\overline{s}(t_2))]$. By (4), there is exactly one $d\in|\mathfrak{A}|$ such that $\vDash_\mathfrak{A} A \ [s(x|d)]$. Hence from (6) and (7) $\overline{s}(t_1)=\overline{s}(t_2)=d$. Thus from (5) and $\nvDash_\mathfrak{A} C \ [s(x|d)]$ and (8) $\vDash_\mathfrak{A} C \ [s(x|d)]$. Contradiction. Hence $\vDash Ix[A, B], \Gamma \Rightarrow \Delta, C_{t_1}$. 
 
\begin{manualtheorem}{\ref{maxconset}}
Any consistent set  of formulas $\Delta$ can be extended to a maximally consistent set $\Delta^+$ such that: 

\noindent (a) for any formula $A$ and variable $x$, if $\neg\forall xA\in \Delta^+$, then for some constant $c$, $\exists !c\in\Delta^+$ and $A_c^x\not\in\Delta^+$; 

\noindent (b) for any formulas $A$ and $B$ and variable $x$, if $Ix[A, B]\in\Delta^+$, then for some constant $c$, $A_c^x, B_c^x\in\Delta^+$ and for all constants $d$, if $A_d^x\in\Delta^+$, then $d=c\in\Delta^+$. 

\noindent (c) for any formulas $A$ and $B$ and variable $x$, if $\neg Ix[A, B]\in\Delta^+$, then for all constants $c$, either $ A_c^x\not\in\Delta^+$, or for some constant $d$, $A_d^x\in\Delta^+$ and $d=c\not\in\Delta^+$, or $B_c^x\not\in\Delta^+$. 
\end{manualtheorem}

\noindent\emph{Proof.} As usual, extend $\mathcal{L}$ to a language $\mathcal{L}^+$ by adding countably new constants ordered by a list $\mathcal{C}=c_1, c_2\ldots$, and extend $\Delta$ by following an enumeration $A_1, A_2\ldots$ of the formulas of $\mathcal{L}^+$ on which every formula occurs infinitely many times as follows:\bigskip

$\Delta_0=\Delta$

\noindent If $\Delta_n, A_n$ is inconsistent, then 

$\Delta_{n+1} = \Delta_n$. 

\noindent If $\Delta_n, A_n$ is consistent, then: 

\noindent (i) If $A_n$ has neither the form $\neg\forall xA$ nor $Ix[A, B]$ nor $\neg Ix[A, B]$, then 

$\Delta_{n+1}=\Delta_n, A_n$ 

\noindent (ii) If $A_n$ has the form $\neg\forall xA$, then 

$\Delta_{n+1}=\Delta_n, \neg\forall xA, \exists ! c, \neg A_c^x$ 

\noindent where $c$ is the first constant of $\mathcal{C}$ that does not occur in $\Delta_n$ or $A_n$; 

\noindent (iii) If $A_n$ has the form $Ix[A, B]$, then 

$\Delta_{n+1}=\Delta_n, Ix[A, B], A_c^x, B_c^x$

\noindent where $c$ is the first constant of $\mathcal{C}$ that does not occur in $\Delta_n$ or $A_n$.  

\noindent (iv) If $A_n$ is has the form $\neg Ix[A, B]$, then

$\Delta_{n+1}=\Delta_n, \neg Ix[A, B], \Sigma_n$

\noindent where $\Sigma_n$ is constructed in the following way. Take a sequence of formulas $\sigma_1, \sigma_2\ldots$ of the form $A_a^x\rightarrow \centerdot \ B_a^x\rightarrow \neg (A_c^x\rightarrow c=a)$, where $a$ is a constant in $\Delta_n, A_n$, and $c$ is a constant on $\mathcal{C}$ not in $\Delta_n, A_n$ or any previous formulas in the sequence. Let $\mathcal{A}=a_1, a_2, \ldots$ be an enumeration of all constants occurring in $\Delta_n, A_n$. In case $\Delta_0$ contains infinitely many formulas, it must be ensured that $\mathcal{C}$ is not depleted of constants needed later. So pick constants from $\mathcal{C}$ by a method that ensures some constants are always left over for later use. The following will do. Let $\sigma_1$ be $A_{a_1}^x\rightarrow \centerdot \ B_{a_1}^x\rightarrow \neg (A_{c_1}^x\rightarrow c_1=a_1)$, where $a_1$ is the first formula of $\mathcal{A}$ and $c_1$ is the first formula of $\mathcal{C}$ not in $\Delta_n, A_n$; let $\sigma_2$ be $A_{a_2}^x\rightarrow \centerdot \ B_{a_2}^x\rightarrow \neg (A_{c_2}^x\rightarrow c_2=a_2)$, where $a_2$ is the second formula on $\mathcal{A}$ and $c_2$ is the $2^2=4$th constant of $\mathcal{C}$ not in $\Delta_n, A_n, \sigma_1$. In general, let $\sigma_n$ be $A_{a_n}^x\rightarrow \centerdot \ B_{a_n}^x\rightarrow \neg (A_{c_n}^x\rightarrow c_n=a_n)$, where $a_n$ is the $n$th constant of $\mathcal{A}$ and $c_n$ is the $2^n$th constant of $\mathcal{C}$ not in $\Delta_n, A_n$ nor any $\sigma_i$, $i<n$. Let the entire collection of $\sigma_i$s be $\Sigma_n$.\bigskip

\noindent $\Delta_{n+1}$ is consistent if $\Delta_n, A_n$ is:\bigskip

\noindent Case (i). Trivial.\bigskip

\noindent Case (ii). Suppose $\Delta_{n+1}=\Delta_n, \neg\forall xA, \exists ! c, \neg A_c^x$ is inconsistent. Then for some finite $\Delta_n'\subseteq\Delta_n$: $\vdash \Delta_n', \neg\forall xA, \exists ! c, \neg A_c^x\Rightarrow\bot$. Hence $\vdash \Delta_n', \neg\forall xA, \exists ! c \Rightarrow A_c^x$ by deductive properties of negation. As $c$ does not occur in any formula in $\Delta_n'$ nor in $\neg\forall xA$, it occurs parametrically, and so by $(R\forall)$, $\vdash\Delta_n', \neg\forall xA\Rightarrow\forall xA$. Hence $\vdash\Delta_n' \Rightarrow\forall xA$, again by deductive properties of negation. But then $\Delta_n', \neg\forall xA$ is inconsistent, and hence so is $\Delta_n, \neg\forall xA$.\bigskip

\noindent Case (iii). Suppose $\Delta_{n+1}=\Delta_n, Ix[A, B], A_c^x, B_c^x$ is inconsistent. Then for some finite $\Delta_n'\subseteq\Delta_n$, $\vdash\Delta_n', Ix[A, B], A_c^x, B_c^x\Rightarrow\bot$. As $c$ does not occur in $\Delta_n', Ix[A, B]$, it occurs parametrically, and hence by $(LI^1)$, $\vdash\Delta_n', Ix[A, B]\Rightarrow\bot$, i.e. $\Delta_n', Ix[A, B]$ is inconsistent, and so is $\Delta_n, Ix[A, B]$.\bigskip 

\noindent Case (iv). Suppose $\Delta_{n+1}=\Delta_n, \neg Ix[A, B], \Sigma_n$ is inconsistent. Then for some finite $\Delta_n'\subseteq\Delta_n$ and a finite $\{\sigma_j\ldots \sigma_k\}\subseteq\Sigma_n$, $\vdash\Delta_n', \neg Ix[A, B], \sigma_j\ldots \sigma_k\Rightarrow\bot$. Let $\sigma_k$ be $A_{a_k}^x\rightarrow \centerdot \ B_{a_k}^x\rightarrow \neg (A_{c_k}^x\rightarrow \ c_k=a_k)$. Then by the deductive properties of implication and negation: 
 
 $\vdash\Delta_n', \neg Ix[A, B], \sigma_j\ldots \sigma_{k-1}\Rightarrow A_{a_k}^x$
 
 $\vdash\Delta_n', \neg Ix[A, B], \sigma_j\ldots \sigma_{k-1}\Rightarrow B_{a_k}^x$ 

 $\vdash\Delta_n', \neg Ix[A, B], \sigma_j\ldots \sigma_{k-1}, A_{c_k}^x\Rightarrow c_k=a_k$
 
\noindent $c_k$ was chosen so as not to occur in any previous $\sigma_i$, $i<k$, nor in $\Delta_n, A_n$. The conditions for $(RI)$ are fulfilled, and so $\vdash\Delta_n', \neg Ix[A, B], \sigma_j\ldots \sigma_{k-1}\Rightarrow Ix[A, B]$. But $\vdash\Delta_n', \neg Ix[A, B], \sigma_j\ldots \sigma_{k-1}\Rightarrow \neg Ix[A, B]$. So $\Delta_n', \neg Ix[A, B], \sigma_j\ldots \sigma_{k-1}$ is inconsistent. Repeat this process from $\sigma_{k-1}$ all the way down to $\sigma_j$, showing that $\Delta_n', \neg Ix[A, B]$ is inconsistent. Hence so is $\Delta_n, \neg Ix[A, B]$.\bigskip 

\noindent Let $\Delta^+$ be the union of all $\Delta_i$. 

$\Delta^+$ is maximal, for if neither $A$ not $\neg A$ are in $\Delta^+$, then there is a $\Delta_k\subseteq\Delta^+$ such that $\Delta_k, A\vdash\bot$ and $\Delta_k, \neg A\vdash\bot$, but then $\Delta_k$ is inconsistent, contradicting the method of construction of $\Delta_k$. 

$\Delta^+$ is consistent, because otherwise some $\Delta_i$ would have to be inconsistent, but they are not. 

$\Delta^+$ satisfies (a) by construction. 

To see that it satisfies (b), suppose $Ix[A, B]\in\Delta^+$. Then there is a $\Delta_{n+1}=\Delta_n, Ix[A, B], A_c^x, B_c^x$, and so $A_c^x, B_c^x\in\Delta^+$. Suppose $A_d^x\in\Delta^+$. Then there is a $\Delta'\subseteq\Delta^+$ such that $\vdash\Delta'\Rightarrow A_c^x$, $\vdash\Delta'\Rightarrow A_d^x$ and by properties of identity $\vdash\Delta'\Rightarrow d=d$. But then by $(LI^2)$, $\vdash\Delta', Ix[A, B] \Rightarrow d=c$, hence $d=c\in\Delta^+$ by the deductive closure of $\Delta^+$. 

To see that it satisfies (c), suppose $\neg Ix[A, B]\in\Delta^+$, but for some constant $c$, $A_c^x\in\Delta$, (1) for all constants $d$, if $A_d^x\in\Delta^+$, then $d=c\in\Delta^+$, and $B_c^x\in\Delta^+$. As every formula occurs infinitely many times on the enumeration of formulas of $\mathcal{L}^+$, there is a $\Delta_n$ that contains $A_c^x$ and $B_c^x$ and $\Delta_{n+1}=\Delta_n, \neg Ix[A, B], \Sigma_n$. Thus $A_c^x\rightarrow \centerdot \ B_c^x \rightarrow \neg (A_b^x\rightarrow b=c)\in\Sigma_n$, for some constant $b$ of $\mathcal{C}$. Consequently, this formula is in $\Delta^+$, too. By the deductive properties of implication and negation and the deductive closure and consistency of $\Delta^+$, (2) $A_b\in\Delta_{n+1}$ and $b=c\not\in\Delta_{n+1}$. But by (1) and (2),  $b=c\in\Delta^+$. Contradiction.\bigskip 

This completes the proof of Theorem \ref{maxconset}. 

\begin{manualtheorem}{\ref{satisfiability}}
If $\Delta$ is a consistent set of formulas, then $\Delta$  is satisfiable. 
\end{manualtheorem}

\noindent \emph{Proof}. Extend $\Delta$ to a maximally consistent set $\Delta^+$ as per Theorem \ref{maxconset}. Next we construct from $\Delta^+$ a structure $\mathfrak{A}$ and function $s\colon V\to |\mathfrak{A}|$. The domain $|\mathfrak{A}|$ of the structure $\mathfrak{A}$ is the set of equivalence classes of terms under identities $t_1=t_2\in\Delta^+$. Denote the equivalence class to which $t$ belongs by $[t]$. The domain of quantification $|\mathfrak{A}^\forall|$ of $\mathfrak{A}$ is the set of equivalence classes of terms $t$ such that $\exists !t\in\Delta^+$.  Clearly $|\mathfrak{A}^\forall|\subseteq|\mathfrak{A}|$. $\mathfrak{A}$ assigns the same set to $\exists!$. Furthermore, $\langle [t_1], \ldots [t_n]\rangle \in P^\mathfrak{A}$ iff $Pt_1\ldots t_n\in\Delta$, $c^\mathfrak{A}=[c]$, for every constant  $c$ (old and new), and $f\mathfrak{A}(t_1\ldots t_n)=[f(t_1\ldots t_n)]$, for any function symbol. For the function $s\colon V\to |\mathfrak{A}|$, $s(x)=[x]$. If follows by induction that $\overline{s}(t)=[t]$. We'll show by induction over the number of connectives in formulas $A$ that\bigskip

\noindent $\vDash_\mathfrak{A} A \ [s]$ if and only if $A\in\Delta^+$.\bigskip

\noindent Suppose $A$ is an atomic formula. (a) $A$ is $Pt_1\ldots t_n$. Then $\vDash_\mathfrak{A} Pt_1\ldots t_n \ [s]$ iff $\langle \overline{s}(t_1)\ldots \overline{s}(t_n)\rangle\in P^\mathfrak{A}$, iff $\langle [t_1]\ldots [t_n]\rangle\in P^\mathfrak{A}$, iff $Pt_1\ldots t_n\in\Delta^+$. (b) $A$ is $t_1=t_2$. Then $\vDash_\mathfrak{A} t_1=t_2 \ [s]$ iff $\overline{s}(t_1)=\overline{s}(t_2)$, iff $[t_1]=[t_2]$, and as these are equivalence classes under identities in $\Delta^+$, iff $t_1=t_2\in\Delta^+$. 

\bigskip

\noindent Suppose $\vDash_\mathfrak{A} A \ [s]$ if and only if $A\in\Delta$, where $A$ has fewer than $n$ connectives.\bigskip

\noindent Case 1. $A$ is $\neg B$. $\vDash_\mathfrak{A} \neg B\ [s]$ iff $\nvDash_\mathfrak{A} B \ [s]$, iff $B\not\in\Delta^+$, by induction hypothesis, iff $\neg B\in\Delta^+$, by maximality of $\Delta^+$.\bigskip

\noindent Case 2. $A$ is $B\rightarrow C$. $\vDash_\mathfrak{A} B\rightarrow C \ [s]$ iff $\nvDash_\mathfrak{A} B$ or $\vDash_\mathfrak{A} C \ [s]$, iff $B\not\in\Delta^+$ or $C\in\Delta^+$, by induction hypothesis, iff $\neg B\in\Delta^+$ or $C\in\Delta^+$, by maximality of $\Delta^+$. $\vdash\Rightarrow\neg B\rightarrow (B\rightarrow C)$ and $\vdash\Rightarrow C\rightarrow (B\rightarrow C)$, so either way, $\Delta^+\vdash B\rightarrow C$, and as $\Delta^+$ is deductively closed, $B\rightarrow C\in\Delta^+$. Conversely, if $B\rightarrow C\in\Delta^+$, then either $\neg B\in\Delta^+$ or $C\in\Delta^+$, as otherwise $B\in\Delta^+$ by maximality, hence $C\in\Delta^+$, by deductive closure, contradiction. So by induction hypothesis, $\nvDash_\mathfrak{A} B$ or $\vDash_\mathfrak{A} C \ [s]$, and thus $\vDash_\mathfrak{A} B\rightarrow C \ [s]$. Hence $\vDash_\mathfrak{A} B\rightarrow C \ [s]$ iff $B\rightarrow C \in\Delta^+$. 
\bigskip

\noindent Case 3. $A$ is $\forall xB$. 

\noindent (a) First, if $\vDash_\mathfrak{A} \forall x B \ [s]$, then $\forall x B\in\Delta^+$. Suppose $\forall xB\not\in\Delta^+$. Then $\neg \forall xB\in\Delta^+$, by maximality, and so for some constant $c$, $\neg B^x_c, \exists !c\in\Delta^+$. Hence by inductive hypothesis $\nvDash_\mathfrak{A} B_c^x \ [s]$ and $\vDash_\mathfrak{A} \exists !c \ [s]$. Hence $c^\mathfrak{A}\in|\mathfrak{A}^\forall|$, and so if $\vDash_\mathfrak{A} \forall xB \ [s]$, then for this $c$, $\vDash_\mathfrak{A} B \ [s(x|\overline{s}(c))]$, so by the Substitution Lemma $\vDash_\mathfrak{A} B^x_c \ [s]$. Contradiction. 

\noindent (b) Next, if $\forall x A\in\Delta^+$, then $\vDash_\mathfrak{A} \forall xA \ [s]$. If $\nvDash_\mathfrak{A} \forall xA \ [s]$, then for some $[t]\in|\mathfrak{A}^\forall|$, $\nvDash_\mathfrak{A} A \ [s(x|[t])]$. By the existence of alphabetic variants, there is a formula $A'$ that is semantically equivalent to $A$ such that $t$ is substitutable for $x$ in $A'$, so $\nvDash_\mathfrak{A} A' \ [s(x|[t])]$.   $\overline{s}(t)=[t]$, so $\nvDash_\mathfrak{A} A' \ [s(x|\overline{s}(t))]$, so by the Substitution Lemma $\nvDash_\mathfrak{A} A'{^x_t} \ [s]$. Hence $A'{^x_t}\not\in\Delta^+$, by induction hypothesis. Hence $\forall xA'\not\in\Delta^+$, by consistency and deductive closure, hence $\forall xA\not\in\Delta^+$, by interdeducibility of alphabetic variants. \bigskip

\noindent Case 4. $A$ is $Ix[A, B]$. 

\noindent (a) First, if $\vDash_\mathfrak{A} Ix[A, B] \ [s]$, then $Ix[A, B]\in\Delta^+$. If $Ix[A, B]\not\in\Delta^+$, then by deductive closure $\neg Ix[A, B]\in\Delta^+$, and so (1) for all constants $c$, either $A_c^x\not\in\Delta^+$, or for some constant $d$, $A_d^x\in\Delta^+$ and $d=c\not\in\Delta^+$, or $B_c^x\not\in\Delta^+$. If $\vDash_\mathfrak{A} Ix[A, B] \ [s]$, then there is a $[t_1]\in|\mathfrak{A}|$ such that: $\vDash_\mathfrak{A} A \ [s(x|[t_1])]$, (2) there is no other $[t_2]\in|\mathfrak{A}|$ such that $\vDash_\mathfrak{A} A \ [s(x|[t_2])]$, and $\vDash_\mathfrak{A} B \ [s(x|[t_1])]$. $\overline{s}(t_1)=[t_1]$, so $\vDash_\mathfrak{A} A \ [s(x|\overline{s}(t_1))]$ and $\vDash_\mathfrak{A} B \ [s(x|\overline{s}(t_1))]$, so by the substitution lemma $\vDash_\mathfrak{A} A_{t_1}^x \ [s]$ and $\vDash_\mathfrak{A} B_{t_1}^x \ [s]$. Thus by induction hypothesis, $A_{t_1}^x, B_{t_1}^x\in\Delta^+$. But by (1) either $A_{t_1}^x\not\in\Delta^+$, or for some constant $d$, $A_d^x\in\Delta^+$ and $d=t_1\not\in\Delta^+$, or $B_{t_1}^x\not\in\Delta^+$. As just established the first and third options are out, which leaves the second, and so by induction hypothesis, $\vDash_\mathfrak{A} A_d^x \ [s]$ and $\nvDash_\mathfrak{A} d=t_1 \ [s]$. So $[d]\not=[t_1]$. By the substitution lemma $\vDash_\mathfrak{A} A\ [s(x|\overline{s}(d))]$, and as $\overline{s}(d)=[d]$, $\vDash_\mathfrak{A} A\ [s(x|[d])]$. But this contradicts (2). Consequently $Ix[A, B]\in\Delta^+$. 

\noindent (b) Next, if $Ix[A, B]\in\Delta^+$, then $\vDash_\mathfrak{A} Ix[A, B] \ [s]$. If $Ix[A, B]\in\Delta^+$, then there is a constant $c$ such that $A_c^x, B_c^x\in\Delta^+$ and for all constants $d$, if $A_d^x\in\Delta^+$, then $d=c\in\Delta^+$. Hence by induction hypothesis $\vDash_\mathfrak{A} A_c^x \ [s]$ and $\vDash_\mathfrak{A} B_c^x \ [s]$. $c^\mathfrak{A}=[c]$, so by the substitution lemma, $\vDash_\mathfrak{A} A \ [s(x|[c])]$ and $\vDash_\mathfrak{A} B \ [s(x|[c])]$. Suppose $\vDash_\mathfrak{A} A \ [s(x|[d])]$, then by induction hypothesis $A_d^x\in\Delta^+$, and so $d=c\in\Delta^+$. Hence $[d]=[c]$ and there is no other $[t]\in|\mathfrak{A}|$ such that $\vDash_\mathfrak{A} A \ [s(x|[t])]$. Hence $\vDash_\mathfrak{A} Ix[A, B] \ [s]$.\bigskip

\noindent Finally, restrict the language again to the language of $\Delta$: structure $\mathfrak{A}$ constructed from $\Delta^+$ satisfies $\Delta$. 

This completes the proof of Theorem \ref{satisfiability}.

\bibliographystyle{splncs04}
\bibliography{KurbisProofSemDefDesc}
\end{document}